\documentclass[a4paper,10pt,twoside]{cpc-hepnp}
\usepackage{CJK,upgreek,fancyhdr}
\usepackage{multicol}
\usepackage{graphicx}
\usepackage{booktabs}
\usepackage{amssymb,bm,mathrsfs,bbm,amscd}
\usepackage[tbtags]{amsmath}
\usepackage{lastpage}

\usepackage{subfig}
\usepackage[pdftex]{hyperref}

\begin{document}
\begin{CJK*}{GB}{gbsn}

\fancyhead[c]{\small ${Chinese\ Physics\ C}$}
\fancyfoot[C]{\small 010201-\thepage}

%\footnotetext[0]{Received 31 June 2015}

\title{A highly pixelated CdZnTe detector based on \textit{Topmetal-${II}^-$} sensor\thanks{Supported by National Natural Science
Foundation of China (11375073, 11305072, U1232206) }}

\author{%
      Shu-Guang Zou(邹曙光)$^{1)}$\email{zoushuguang@mails.ccnu.edu.cn}%
	\quad Yan Fan(樊艳)
	\quad Xiang-Ming Sun(孙向明)$^{2)}$\email{sphy2007@126.com}
	\quad Guang-Ming Huang(黄光明)$^{3)}$\email{gmhuang@phy.ccnu.edu.cn}\\
	\quad Hua Pei(裴骅)
	\quad Zhen Wang (王珍)
	\quad Jun Liu (刘军)
	\quad Ping Yang (杨苹)
	\quad Dong Wang (王东)
}
\maketitle

\address{%
PLAC, Key Laboratory of Quark $\&$ Lepton Physics (MOE), Central China Normal University, Wuhan, Hubei 430079, P.R.China, Grant No.:QLPL2015P01
}

\begin{abstract}
\textit{Topmetal-${II}^-$} is a low noise CMOS pixel direct charge sensor with a pitch of 83$\mu m$. CdZnTe is an excellent semiconductor material for radiation detection. The combination of CdZnTe and the sensor makes it possible to build a detector with high spatial resolution. In our experiments, an epoxy adhesive is used as the conductive medium to connect the sensor and Cadmium Zinc Telluride (CdZnTe). The diffusion coefficient and charge efficiency of electrons are measured at a low bias voltage of -2 Volts, and the image of a single alpha is clear with a reasonable spatial resolution. The detector of such structure has the potential to be applied in X-ray imaging systems with a further improvements of the sensor. 
\end{abstract}



%\footnotetext[0]{\hspace*{-3mm}\raisebox{0.3ex}{$\scriptstyle\copyright$}2013
%Chinese Physical Society and the Institute of High Energy Physics
%of the Chinese Academy of Sciences and the Institute
%of Modern Physics of the Chinese Academy of Sciences and IOP Publishing Ltd}%

\begin{multicols}{2}

\section{Introduction}
CdZnTe is an excellent semiconductor material for radiation detection with high resistivity, large atomic number and high band gap. CdZnTe detector has been developed for the last twenty years. CdZnTe detectors with a high spatial resolution have great potential to be applied in applications such as Compton camera\cite{compton_camera}, medical imaging system and neutrinoless double beta decay experiments\cite{CdZnTe_neutrino}. 

CdZnTe and readout module are commonly connected by metal electrodes. In general, they can achieve high energy resolution\cite{energyResolution1,energyResolution2}. CdZnTe detectors with different structure of single crystal were developed to improve spatial resolution\cite{virtual_frisch,coplanar_grid} for applications. And in other way, pixelated CdZnTe detectors were developed for improving the spatial resolution. A spatial resolution of hundreds of $\mu$m have been achieved to detect X-ray\cite{pixel_CdZnTe_2001,highly_pixelated,144channel_readout}. For small pixel detectors of CdZnTe, signal induced by a single particle will be shared by multiple pixels, and charge sharing effect have been studied in detail\cite{chargeShare,czt_diffusion}. Therefore, in this paper, we will not consider those effects.

\textit{Topmetal-${II}^-$} is a highly pixelated direct charge sensor with a rather low noise and high spatial resolution. It can be applied into a \textbf{Time Projection Chamber (TPC)} as a charge collector to measure single electrons generated by alpha particles\cite{topmetal} at room temperature without any charge multiplier being necessary. This result prompted us to test if \textit{Topmetal-${II}^-$} can be coupled with CdZnTe to measure signals radiated by alpha particles at room temperature.

\section{Experiment setup}

%	\subsection{description of system}
\textit{Topmetal-${II}^-$} is a low noise CMOS pixel direct charge sensor that contains a 72$\times$72 pixel array of 83 ${\mu}m$ pitch size with a sensitive area of about 6$\times$6 mm$^2$.  The size of CdZnTe is about 5.8$\times$5.8$\times$2 mm$^3$, which is slightly smaller than the sensitive area of \textit{Topmetal-${II}^-$} sensor for protecting the bonding wires. When the sensor and the CdZnTe are combined together, the pixel size of determined by the pixel size of the sensor.

\begin{center}
	\centering
	\includegraphics[width=0.8\columnwidth]{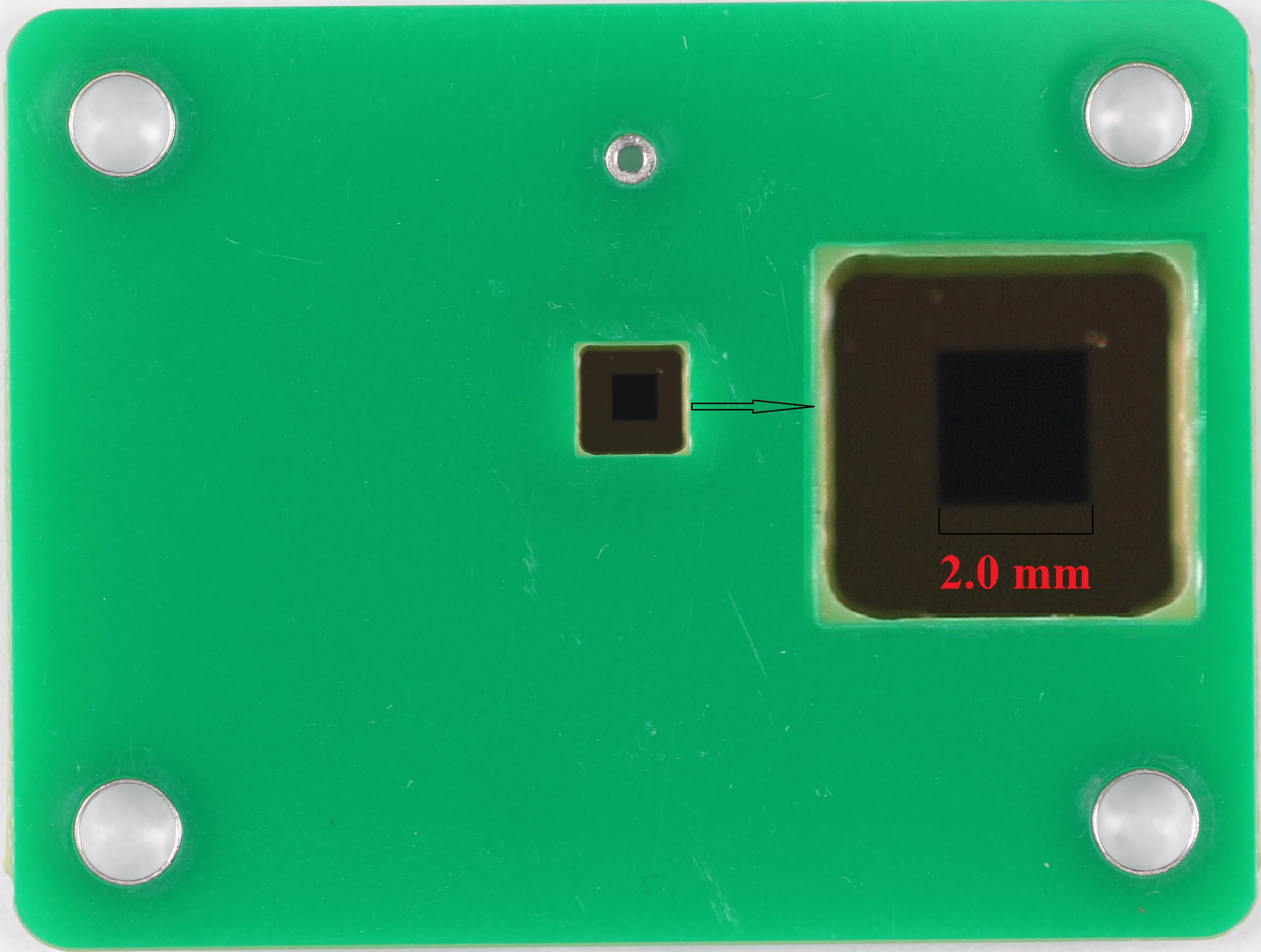}
	\figcaption{\label{CdZnTe with PCB board}The side of a CdZnTe crystal plated with an electrode is coupled to a PCB board. The size of CdZnTe is 5.8$\times$5.8$\times$2 mm$^3$. The little square region (black) in the center of the crystal that is deliberately naked without an electrode is about 2$\times$2 mm$^2$.}
\end{center}

%	\subsection{mechanical structure}
Only one side of the CdZnTe crystal is plated with a metal electrode (gold) fixed on a Printed Circuit Board (PCB) with a conductive adhesive (3M Scotch-Weld$^{TM}$ Epoxy Potting Compound DP270 Clear, 3M Company, USA) shown in Figure \ref{CdZnTe with PCB board}, and the inner open square is deliberately naked without an electrode for signal injection. The other side without a metal electrode totally is used to connect with \textit{Topmetal-${II}^-$} using an epoxy adhesive. The CdZnTe-\textit{Topmetal-${II}^-$} structure is based on a PCB board as shown in the bottom of Figure \ref{structure:a}, which is fixed on a mechanical plane that can be moved up and down or left and right. 

\begin{center}
	\centering
	\captionsetup{type=figure}
	\subfloat[]{\includegraphics[width=0.85\columnwidth]{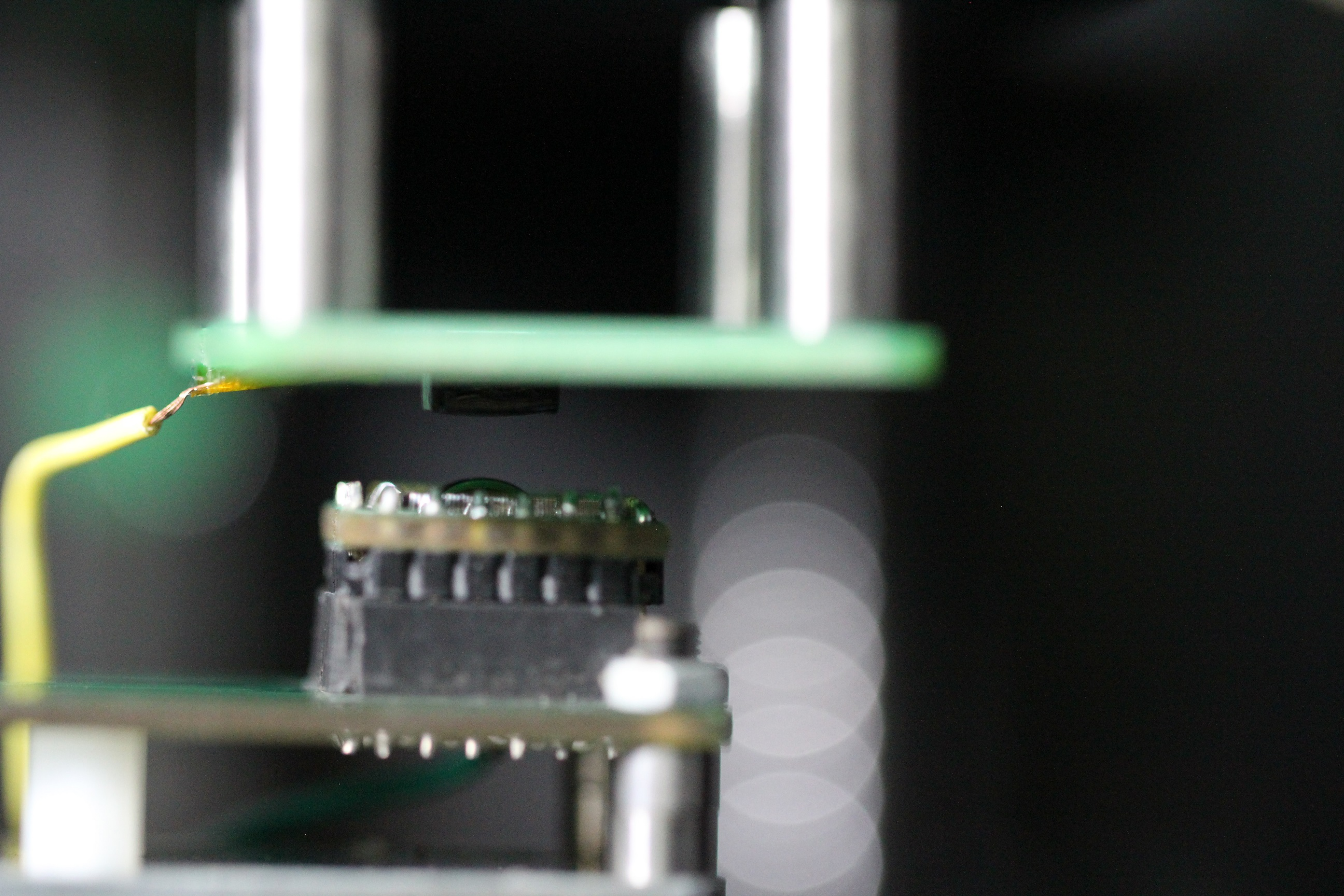}\label{structure:a}}
	\hfill
	\subfloat[]{\includegraphics[width=0.95\columnwidth]{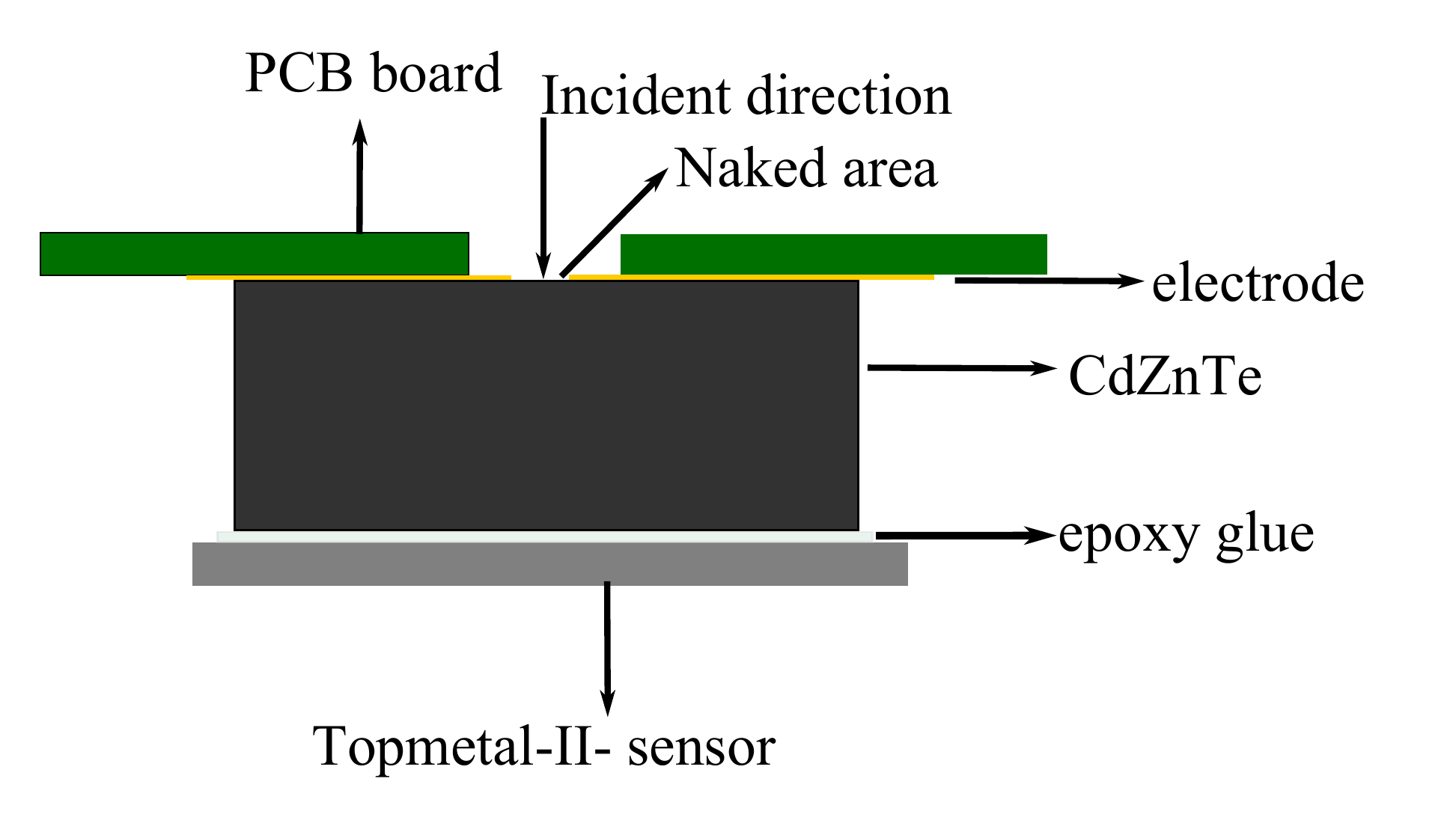}\label{structure:b}}
	\figcaption{Only one side of the crystal is plated with a metal electrode. The other side that connected to \textit{Topmetal-${II}^-$} does not have a metal electrode, so the connection between the crystal and the sensor is certainly not ohmic contact. (a) Photograph of experiments setup. It consists of \textit{Topmetal-${II}^-$} sensor, a CdZnTe crystal and readout system(not shown). (b) schematic diagram of experiments setup. CdZnTe crystal is coupled to \textit{Topmetal-${II}^-$} sensor by an epoxy adhesive.}
\end{center}

First, the epoxy adhesive is dropped onto the \textit{Topmetal-${II}^-$} sensor by a syringe manually, and then the sensor is inserted into the slot on the PCB board that is fixed on the mechanical plane. As shown in Figure \ref{structure:a}, we can move the crystal slowly down to the sensor, and keep the sensor parallel to the crystal appropriately. Figure \ref{structure:b} shows the basic principle structure of the CdZnTe detector. Using an epoxy adhesive as the medium to connect the \textit{Topmetal-${II}^-$} sensor and CdZnTe crystal is a way more simple and convenient compared to bump bonding.

\section{Results}

We mainly test the noise, diffusion, and efficiency of the detector system.
\subsection{Noise Test}
The equivalent noise charge (ENC) of \textit{Topmetal-${II}^-$} is about 13 e${^-}$\cite{noise} with no CdZnTe connected under the condition of V$_reset$=800mV and V$_ref$=618mV\cite{topmetal,noise}. When CdZnTe is connected to \textit{Topmetal-${II}^-$} via the epoxy adhesive, the electronic noise of baseline of the same pixels increases about 3 times (from 2.2 mV to 8.3 mV) with a bias voltage -2V of CdZnTe, which is quite bigger.
\subsection{Diffusion}	

\begin{center}
	\captionsetup{type=figure}
	\centering
	\includegraphics[width=1.0\columnwidth]{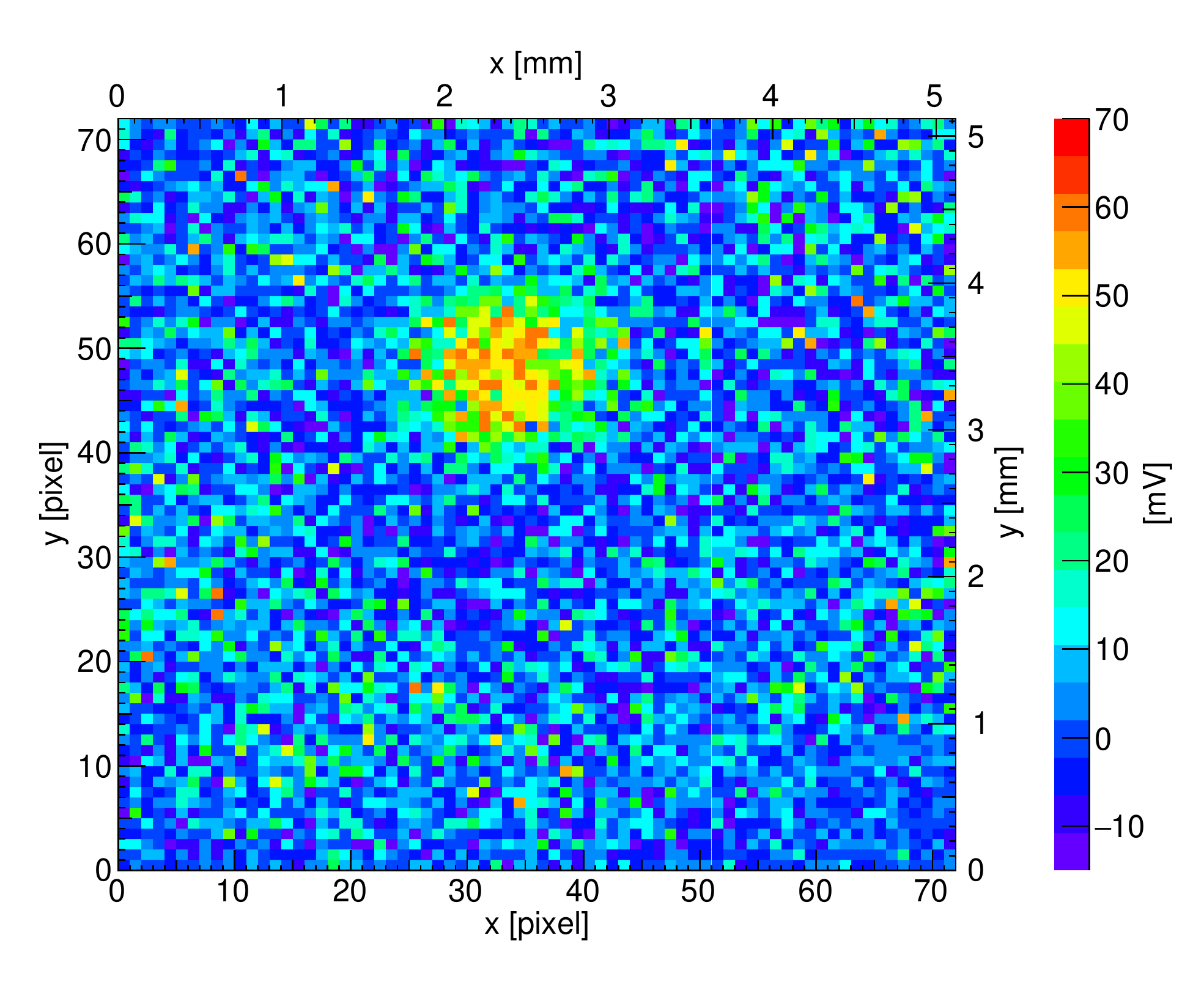}
	\centering
	\figcaption{A 650 nm laser image by CdZnTe and \textit{Topmetal-${II}^-$} sensor. The expanded diameter of the induced charge cloud by a 650 $nm$ laser is about 420 ${\mu}m$ at a bias of -2 Volts.}	
	\label{diffusion image}		
\end{center}

There is a 2$\times$2 mm$^2$ area without a plated electrode on the surface of CdZnTe crystal, through which a 650 nm laser signal can be injected. The bias voltage supplied to the CdZnTe is about -2 Volts. Because the gain of the pre-amplifier of \textit{Topmetal-${II}^-$} is high, the back bias voltage of the crystal should be low to avoid the analog output saturation. We have observed that the expanded diameter of the induced charge cloud by a 650 $nm$ laser is about 420 ${\mu}m$ \textbf{(about 5 pixels)} at a bias of -2 Volts shown in Figure \ref{diffusion image}. In theory, the electron cloud expands to 318 ${\mu}m$ after drifting through the entire thickness of the crystal\cite{czt_diffusion}. The expanded diameter we observed is larger than the theoretical calculation taking no account of charges diffusion during the epoxy adhesive. The medium of the epoxy adhesive is an important reason for this result.

\subsection{Efficiency}

\begin{center}
	\centering
	\captionsetup{type=figure}
	\subfloat[]{\includegraphics[width=1.0\columnwidth]{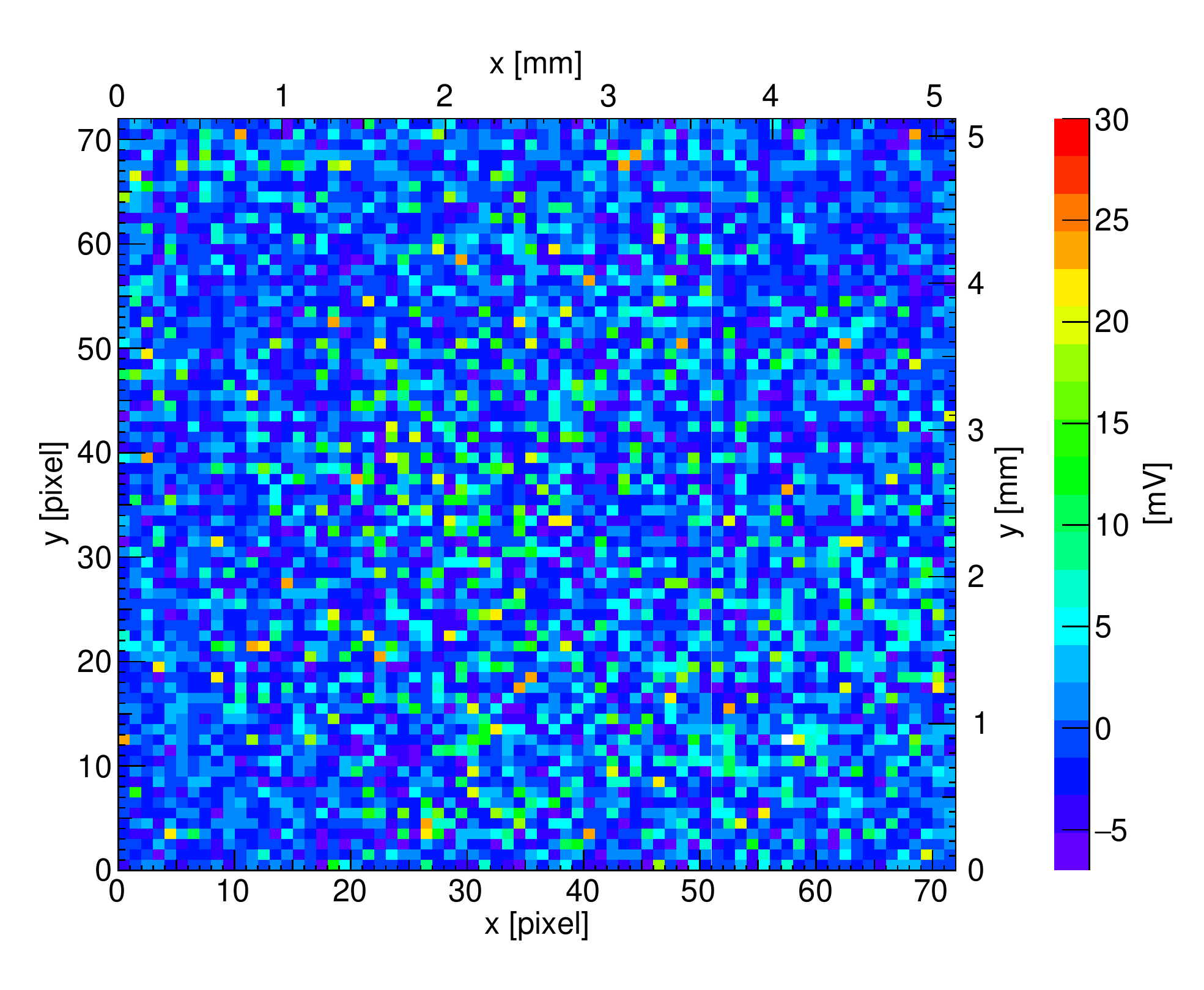}\label{alaph:pede}}
	\hfill
	\subfloat[]{\includegraphics[width=1.0\columnwidth]{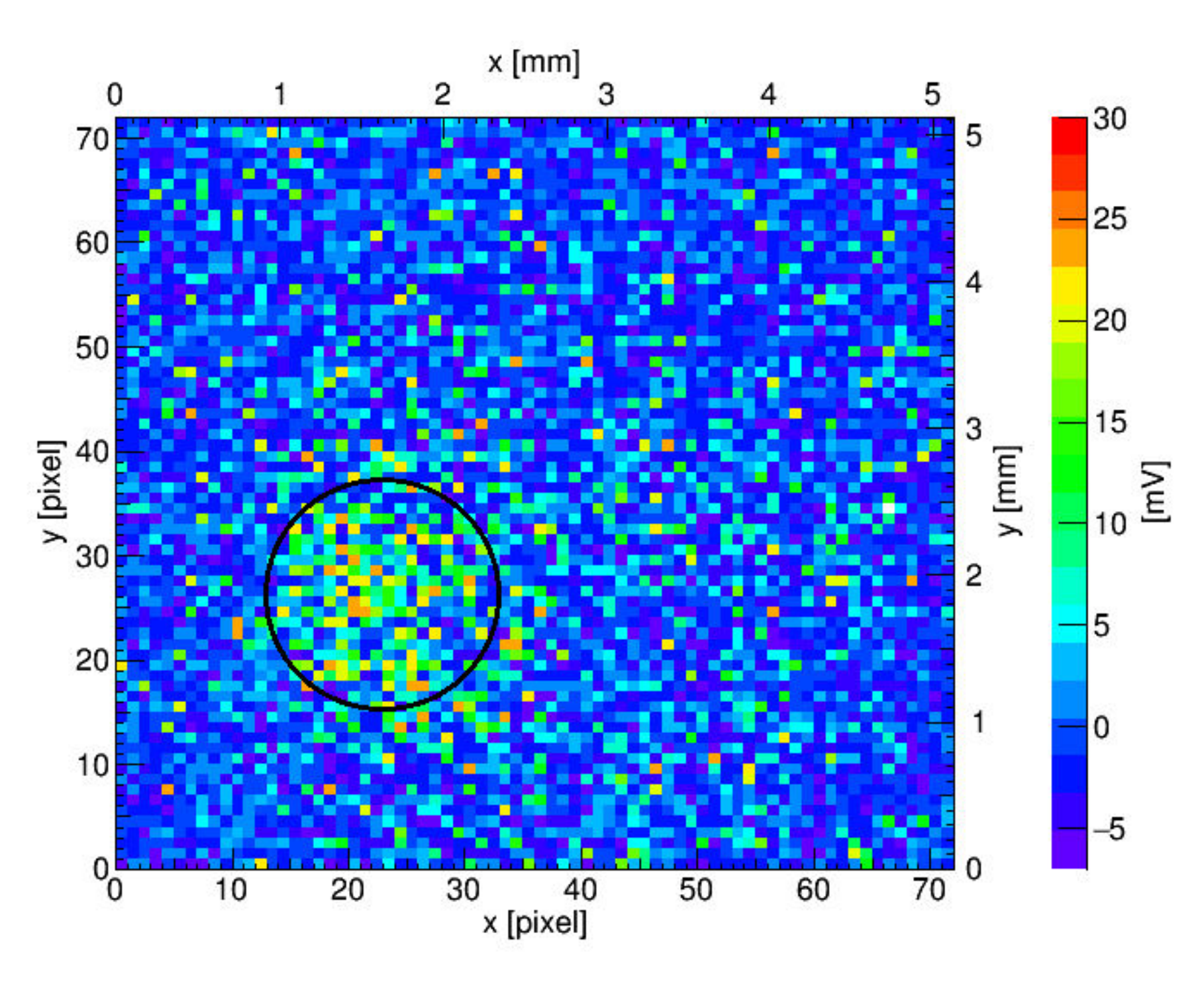}\label{alpha:alpha}}
	\figcaption{(a) Two dimensional image of \textit{Topmetal-${II}^-$} without particles injected into CdZnTe (b) Alpha image by CdZnTe and \textit{Topmetal-${II}^-$} sensor. The alpha source is placed on the top of the CdZnTe crystal. Alpha particles can pass through the air between CdZnTe and plane of alpha source and inject into CdZnTe. The charge cloud generated by alphas will be collected by \textit{Topmetal-${II}^-$} sensor to form a two dimensional image. The expanded diameter of the induced charge cloud by a alpha particle is about 500 ${\mu}m$ (about 6 pixels) at a bias of -2 Volts.}
\end{center}

To calculate the efficiency of the detector, we use a collimated alpha source to inject signal into the CdZnTe crystal. The number of particles detected by a Geiger counter is about 300 per minute. For comparison, we put a image of the sensor without any signal injected in Figure \ref{alaph:pede}.

The collimated alpha source is placed on the top of the crystal corresponding to the part without an electrode (Figure \ref{structure:b} Naked area). The size of the collimated hole that alpha particles can pass through is about 1.5 mm. The image of a single alpha is clear with a reasonable spatial resolution shown in Figure \ref{alpha:alpha} and the counts are comparable to Geiger counter. We can loop pixels to sum the value of signals generated by alpha particles to get the total charges value. The saturated pixels are not revised and treated as bad pixels. The alpha energy is about 5.4 MeV. The ionization $W$-value of CdZnTe is about 4.6 eV. The charge collection efficiency of CdZnTe detector coupled with \textit{Topmetal-${II}^-$} is estimated $\sim$3.5\%, which is rather low.

\section{Summary}
\textit{Topmetal-${II}^-$} is a low noise CMOS pixel direct charge sensor that contains a 72$\times$72 pixel array of 83 ${\mu}m$ pitch size. CdZnTe is an excellent semiconductor material for radiation detection. We use an epoxy adhesive as the media to connect the \textit{Topmetal-${II}^-$} sensor to CdZnTe crystal to build a detector that has a good spatial resolution concerning the -2 volts bias voltage. Besides this method is more simple and convenient compared to bump bonding. Then we test the performance of such an CdZnTe detector coupled with \textit{Topmetal-${II}^-$} sensor. We have observed that the expanded diameter of the induced charge cloud by a 650 $nm$ laser is about 420 ${\mu}m$ at a back bias of -2 Volts, and the image of a single alpha is clear with a reasonable spatial resolution as the charge cloud induced by 650 laser. However, at such a low voltage, the charge collection efficiency of CdZnTe detector coupled with \textit{Topmetal-${II}^-$} is very low ($\sim$3.5\%). 

\section{Outlook}
Through the experiment, we find that it is feasible to use this simple method to get a CdZnTe detector with a reasonable spatial resolution corresponding to the bias voltage. Due to the characteristics of the sensor itself, we can not supply a higher bias voltage (such as 10 V) on the same crystal, the diffusion of charge signal on the crystal is very large and the spatial resolution is not very high.  In order to get a good enough spatial resolution that matches to the size of pixels, we need to continue improving the bias voltage to minimize diffusion of charge inside the CdZnTe crystal by optimizing the design of the sensor and obtaining CdZnTe crystals with a higher resistance.

\acknowledgments{Thank the PLAC group who contributed to the \textit{Topmetal-${II}^-$} sensor design and test. Besides we are grateful to the Imdetek company for providing us the CdZnTe crystals. }

\end{multicols}
\begin{multicols}{2}

\end{multicols}

\clearpage
\end{CJK*}
\end{document}